\documentclass[
    aps,                 
    prd,                 
    reprint,             
    superscriptaddress,  
    nofootinbib,         
    eprint,              
]{revtex4-2}

\usepackage{graphicx}   
\usepackage{subfigure}  
\usepackage{dcolumn}    
\usepackage{bm}         
\usepackage{multirow}   

\usepackage{amsmath}
\usepackage{amssymb}
\usepackage{amsthm}
\usepackage{mathrsfs}
\usepackage{amsfonts}

\usepackage{hyperref}
\hypersetup{
    colorlinks=true,    
	linkcolor=blue,     
	citecolor=green,    
	filecolor=magenta,  
	urlcolor=cyan,      
}

\usepackage{cleveref}
\makeatletter
\AddToHook{cmd/appendix/before}{\def\cref@section@alias{appendix}}  
\makeatother

\usepackage{glossaries-extra}  
\glsdisablehyper               
\newglossaryentry{BH}{
    name=BH,
    description=black hole,
    first=black hole (BH),
    plural=BHs,
    descriptionplural=black holes,
    firstplural=black holes (BHs)
}

\newglossaryentry{BS}{
    name=BS,
    description=boson star,
    first=boson btar (BS),
    plural=BSs,
    descriptionplural=boson stars,
    firstplural=boson stars (BSs)
}

\newglossaryentry{GR}{
    name=GR,
    description=General Relativity,
    first=General Relativity (GR)
}

\newglossaryentry{EoM}{
    name=EoM,
    description=equation of motion,
    first=equation of motion (EoM),
    plural=EoM,
    descriptionplural=equations of motion,
    firstplural=equations of motion (EoM)
}

\newglossaryentry{GW}{
    name=GW,
    description=gravitational wave,
    first=gravitational wave (GW),
    plural=GWs,
    descriptionplural=gravitational waves,
    firstplural=gravitational waves (GWs)
}

\newglossaryentry{LISA}{
    name=LISA,
    description=Laser Interferometer Space Antenna,
    first=Laser Interferometer Space Antenna (LISA)
}

\newglossaryentry{EMS}{
    name=EMS,
    description=Einstein-Maxwell-Scalar,
    first=Einstein-Maxwell-Scalar (EMS)
}

\newglossaryentry{ADM}{
    name=ADM,
    description=Arnowitt-Deser-Misner,
    first=Arnowitt-Deser-Misner (ADM)
}

\newglossaryentry{BSSN}{
    name=BSSN,
    description=Baumgarte-Shapiro-Shibata-Nakamura,
    first=Baumgarte-Shapiro-Shibata-Nakamura (BSSN)
}

\newglossaryentry{CTT}{
    name=CTT,
    description=conformal transverse-traceless,
    first=conformal transverse-traceless (CTT)
}

\newglossaryentry{EM}{
    name=EM,
    description=Einstein-Maxwell,
    first=Einstein-Maxwell (EM)
}

\newglossaryentry{RN}{
    name=RN,
    description=Reissner–Nordström,
    first=Reissner–Nordström (RN)
}

\begin{document}

\title{Electromagnetic duality degeneracy in dynamical black hole mergers}


\author{José Ferreira}
\email{jpmferreira@ua.pt}
\affiliation{Departamento de Matemática da Universidade de Aveiro and Centre for Research and Development in Mathematics and Applications (CIDMA), Campus de Santiago, 3810-193 Aveiro, Portugal}

\author{Gabriele Bozzola}
\email{gbozzola@amazon.com}
\thanks{Work done prior to joining AWS}
\affiliation{AWS Center for Quantum Computing, Pasadena, 91125}

\author{Carlos A. R. Herdeiro}
\email{herdeiro@ua.pt}
\affiliation{Departamento de Matemática da Universidade de Aveiro and Centre for Research and Development in Mathematics and Applications (CIDMA), Campus de Santiago, 3810-193 Aveiro, Portugal}

\author{Vasileios Paschalidis}
\email{vpaschal@arizona.edu}
\affiliation{Department of Astronomy, University of Arizona, Tucson, AZ, USA}
\affiliation{Department of Physics, University of Arizona, Tucson, AZ, USA}

\author{Miguel Zilhão}
\email{mzilhao@ua.pt}
\affiliation{Departamento de Física da Universidade de Aveiro and Centre for Research and Development in Mathematics and Applications (CIDMA), Campus de Santiago, 3810-193 Aveiro, Portugal}

\date{May 2026}

\begin{abstract}
  Electromagnetic duality is a symmetry of the source-free Einstein–Maxwell equations that rotates electric and magnetic fields while leaving the stress–energy tensor invariant. We present the first fully nonlinear realization of this symmetry in dynamical strong-gravity regimes by performing numerical relativity simulations of charged black hole mergers across a continuous duality family. Starting from electrically charged binaries, we generate dyonic and magnetically charged configurations via duality rotations and evolve them within a common numerical framework. We find that all dual configurations exhibit identical spacetime dynamics, while the emitted electromagnetic radiation is related by a rotation of its polarization equal to the duality angle. Our results demonstrate a degeneracy of gravitational observables under electromagnetic duality and provide a concrete mapping between dual configurations at the level of radiation, establishing electromagnetic duality as an organizing principle for dynamical Einstein–Maxwell solutions.
\end{abstract}

\maketitle


\section{Introduction}
\label{sec:introduction}

Electromagnetism possesses a remarkable symmetry in the absence of sources: the electric and magnetic fields can be continuously rotated into one another without altering Maxwell’s equations \cite{Jackson:1998nia,Landau:1987gn}. This electric--magnetic duality leaves the stress--energy tensor invariant and therefore preserves the spacetime geometry in the \gls{EM} theory. As a consequence, every solution of the \gls{EM} equations belongs to a continuous family of dual configurations that share the same spacetime structure but differ in their electromagnetic field content.

This symmetry has long been recognized in classical field theory and plays an important role in several areas of theoretical physics, ranging from classical electromagnetism to quantum field theory and string theory, e.g.,~\cite{Deser:1976iy,Gaillard:1981rj}. In the context of General Relativity and some of its extensions, electromagnetic duality implies that purely electrically charged, purely magnetically charged, and dyonic configurations can be related by simple rotations of the electromagnetic field \cite{Misner:1973prb,Stephani:2003tm} and it serves as a solution generating technique; see, e.g., \cite{Gibbons:1995cv,Ortin:2015hya,Herdeiro:2020iyi,Bokulic:2025usc}. Related ideas have also been explored for gravitational perturbations, where electric--magnetic duality at the light ring has been argued to organize eikonal quasinormal-mode isospectrality in effective field theories~\cite{Bah:2026aia}. For stationary solutions, such as the \gls{RN} \gls{BH}, this symmetry is well understood: the spacetime metric depends only on the invariant combination of electric and magnetic charges, while the electromagnetic field rotates within the duality family.

Much less attention has been given to the role of electromagnetic duality in fully dynamical spacetimes. In particular, in systems involving strong gravitational dynamics -- such as binary \gls{BH} mergers -- the interplay between electromagnetic fields and spacetime curvature raises natural questions. Do members of the same duality family remain dynamically indistinguishable at the level of the evolving spacetime geometry? If so, how does the duality manifest itself in observable electromagnetic radiation emitted by such systems?

These questions are especially relevant in the context of numerical relativity simulations of the \gls{EM} system. Over the past decades, significant progress has been made in evolving charged \gls{BH} spacetimes numerically, including studies of binary mergers and their associated electromagnetic emission \cite{Alcubierre2009,Zilhao2012, Zilhao:2013nda,Bozzola2019,Bozzola2020,Bozzola2021,Bozzola2022,Luna2022,Bozzola2023,Smith2024,Smith2024a} and non-linear stability studies \cite{Zilhao2014}. In these simulations, the electromagnetic sector is typically initialized with purely electric charges. However, electromagnetic duality suggests that such configurations represent only one member of a broader family of solutions that includes dyonic and magnetically charged systems. Understanding how this symmetry manifests itself in dynamical evolutions is therefore both conceptually interesting and practically useful.

In this work we explore electromagnetic duality in dynamical \gls{EM} spacetimes through fully nonlinear numerical simulations of binary \gls{BH} mergers. Starting from initial data describing a pair of electrically charged \glspl{BH}, we generate a continuous family of dual configurations by applying the standard duality rotation to the electromagnetic field. These configurations correspond to binaries with different electric-magnetic compositions of charge, including purely electric, dyonic, and purely magnetic. Since the stress-energy tensor is invariant under the duality transformation, the spacetime geometry is expected to evolve identically across the entire dual family. We remark that obtaining initial data for a binary of dyonic BHs is nontrivial and an open problem, except at the points along a duality ``orbit" that includes purely electrically charged BHs, where one can make use of the approach and code reported in~\cite{Bozzola2019}.

Our numerical evolutions confirm this expectation. We find that the spacetime dynamics -- such as the trajectories of the \glspl{BH} and the merger time -- are indistinguishable for all configurations related by duality rotations. At the same time, the electromagnetic radiation emitted by the system carries a clear imprint of the duality transformation. In particular, the polarization of the outgoing electromagnetic waves is rotated by an angle equal to the duality parameter, providing a direct observational signature that distinguishes different members of the same dual family.

These results highlight a duality degeneracy in dynamical \gls{EM} systems: gravitational observables are insensitive to the electric-magnetic composition of the charges, while electromagnetic radiation encodes the duality angle through its polarization structure. From a practical perspective, this symmetry also provides a straightforward method to generate dynamical dyonic or magnetically charged solutions directly from simulations of electrically charged systems, without modifying the underlying numerical infrastructure. Conversely, these simulations provide a check on the infrastructure for evolving systems with magnetic charge.

The remainder of this paper is organized as follows. In \cref{sec:duality} we review the electromagnetic duality in the \gls{EM} system and describe how dual solutions can be constructed from a given electromagnetic configuration. In \cref{sec:evolutions} we present fully nonlinear numerical evolutions of binary \gls{BH} mergers across a duality family and analyze the resulting electromagnetic radiation. Finally, in \cref{sec:final-remarks} we summarize our results and discuss their implications.

\section{Electromagnetic Duality}
\label{sec:duality}

We consider the \gls{EM} model in four spacetime dimensions. The action for this model is%
\footnote{Throughout this work we consider geometrized units ($G=c=1$), use Greek indices to represent 4D quantities and Latin indices represent 3D quantities.}
\begin{equation}
\mathcal S = \frac{1}{16 \pi}   \int \sqrt{-g} \, \left({R} - F_{\mu \nu}F^{\mu \nu} \right) d^4x \,,
\end{equation}
where $g$ is the determinant of the metric $g_{\mu\nu}$, $R$ is the Ricci scalar, and $F_{\mu\nu}= \nabla_\mu A_\nu - \nabla_\nu A_\mu$ is the Faraday tensor with $A_\mu$ the 4-potential. Variation with respect to the metric yields
\begin{equation}
    \label{eq:EFE}
    G_{\mu\nu} = 8\pi\,T_{\mu\nu} \,,
\end{equation}
with the stress-energy tensor
\begin{equation}
    \label{eq:Tmunu}
    T_{\mu\nu}=\frac{1}{4\pi}\left(F_{\mu\alpha}F_{\nu}{}^{\alpha} -\frac{1}{4}g_{\mu\nu}F_{\alpha\beta}F^{\alpha\beta}\right) \,.
\end{equation}
The covariant Maxwell equations read
\begin{subequations}
\label{eq:EoM-Fmunu}
\begin{align}
    \nabla_\mu F^{\mu\nu}             &= 0 \,, \\
    \nabla_\mu {}^{\star}\!F^{\mu\nu} &= 0 \,,
\end{align}
\end{subequations}
where ${}^{\star}\!F_{\mu\nu}\equiv-\tfrac12\,\epsilon_{\mu\nu\rho\sigma}F^{\rho\sigma}$ is the Hodge dual ($\epsilon_{\mu\nu\rho\sigma}$ is the Levi-Civita tensor with the convention $\epsilon_{1230} = \sqrt{-g}$). Given the absence of sources, Maxwell's equations are known to be invariant under the continuous duality rotation by an angle $\alpha$
\begin{subequations}
\label{eq:duality}
\begin{align}
    \hat{F}_{\mu \nu} &= F_{\mu \nu} \cos\alpha + {}^{\star}\!F_{\mu \nu}\sin\alpha \,, \\
    {}^{\star}\!\hat{F}_{\mu\nu} &= {}^{\star}\!F_{\mu\nu} \cos\alpha - F_{\mu \nu}\sin\alpha \,,
\end{align}
\end{subequations}
where the hat denotes the transformed quantity. This rotation also leaves the stress-energy tensor \cref{eq:Tmunu} unchanged.

Following an Eulerian observer that is moving with a 4-velocity $n^\mu$, we can write the Faraday tensor and its Hodge dual with the more familiar electric and magnetic fields
\begin{subequations}
\label{eq:Decomposition-Fmunu}
\begin{align}
    F_{\mu \nu} = n_\mu E_\nu - n_\nu E_\mu + \epsilon_{\mu\nu\alpha\beta} n^\beta B^\alpha \,, \\
    {}^{\star}\!F_{\mu\nu} = n_\mu B_\nu - n_\nu B_\mu - \epsilon_{\mu\nu\alpha\beta} n^\beta E^\alpha \,,
\end{align}
\end{subequations}
and applying \cref{eq:duality} we obtain the transformation
\begin{subequations}
\label{eq:duality-fields}
\begin{align}
    \hat{E}_\mu = E_\mu \cos \alpha + B_\mu \sin \alpha \,, \\
    \hat{B}_\mu = B_\mu \cos \alpha - E_\mu \sin \alpha \,.
\end{align}
\end{subequations}

Therefore, for any solution of the \gls{EM} model with $(E_\mu, B_\mu)$, there is a family of dual solutions $(\hat{E}_\mu, \hat{B}_\mu)$ that are related by the duality \cref{eq:duality-fields} with parameter $\alpha$, whose dynamics is expected to be the same. For dynamical spacetimes it is sufficient for two configurations to be dual to each other at a single point in time (say for the initial data), for their evolution to be duality equivalent.

The implication is that any numerical or analytical construction of initial data for the \gls{EM} system, in electrovacuum, can be straightforwardly extended by applying the duality transformation \cref{eq:duality-fields} to generate new initial data and new evolutions, albeit in the same duality ``orbit".

\subsection{Dual Solutions}
\label{subsec:dual-solutions}

Our interest is studying the duality in dynamical spacetimes. Perhaps the simplest dynamical solutions known for \gls{EM} are those of a binary system of charged \glspl{BH}. Let us assume a solution that, at a given point in time, can be described by two electrically charged point charges. The electric and magnetic fields are given by%
\footnote{As written, these fields are divergenceless in flat spacetime. In the initial data construction, the same Coulomb profile is adopted for the conformally rescaled fields, which satisfy a flat-space divergence-free condition in conformal space (see \cref{app:initial-data}). The present expressions therefore represent these conformal fields, and the charge transformation under duality \cref{eq:duality-fields} applies without modification.}
\begin{subequations}
\label{eq:binary-electromagnetic}
\begin{align}
    E^i &= Q_1 \frac{\left(\vec{r}-\vec{r}_1\right)^i}{|\vec{r}-\vec{r}_1|^3} + Q_2 \frac{\left(\vec{r}-\vec{r}_2\right)^i}{|\vec{r}-\vec{r}_2|^3} \,, \\
    B^i &= 0 \,,
\end{align}
\end{subequations}
where $(M_n,Q_n,\vec{r}_n)$ are the mass, electric charge and position of the $n$-th \gls{BH}. Applying \cref{eq:duality-fields} we obtain
\begin{subequations}
\begin{align}
    \hat{E}^i &= \hat{Q}_1 \frac{\left(\vec{r}-\vec{r}_1\right)^i}{|\vec{r}-\vec{r}_1|^3} + \hat{Q}_2 \frac{\left(\vec{r}-\vec{r}_2\right)^i}{|\vec{r}-\vec{r}_2|^3} \,, \\
    \hat{B}^i &= \hat{P}_1 \frac{\left(\vec{r}-\vec{r}_1\right)^i}{|\vec{r}-\vec{r}_1|^3} + \hat{P}_2 \frac{\left(\vec{r}-\vec{r}_2\right)^i}{|\vec{r}-\vec{r}_2|^3} \,,
\end{align}
\end{subequations}
where $(\hat{Q}_n,\hat{P}_n)$ are the dual electric and magnetic charges, given by
\begin{subequations}
\begin{align}
    \hat{Q}_n &=  Q_n \cos \alpha \,, \\
    \hat{P}_n &= -Q_n \sin \alpha \,.
\end{align}
\end{subequations}
Therefore, the dual solution describes two electric, dyonic or magnetically charged \glspl{BH}, depending on the value of $\alpha$, with the same mass and position as the original ones.

\begin{figure}
  \centering
  \includegraphics[width=0.75\linewidth]{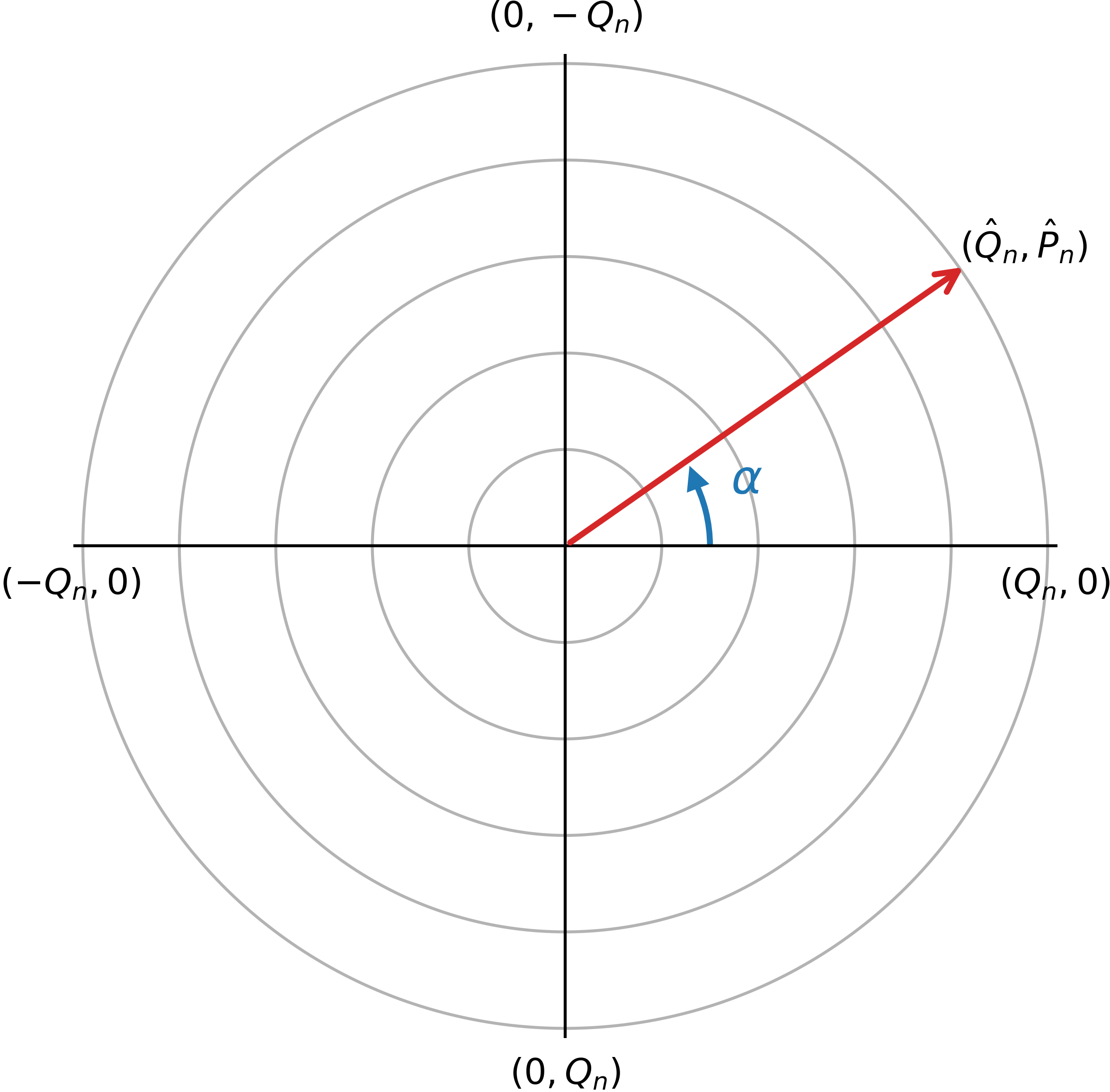}
  \caption{Family of solutions labeled by angle $\alpha$ that are dual to the binary of electrically charged \glspl{BH}, with $\hat{Q}_i$ representing dual electric charge and $\hat{P}_i$ the dual magnetic charge.}
  \label{fig:dual-to-Q-angle}
\end{figure}

Thus, from an electromagnetic point of view, we characterize the solutions via the charges and angle $(Q_n,\alpha)$. From a gravitational point of view, we characterize the spacetime via the mass and charges $(M_n,Q_n)$. Therefore, solutions with equal $(M_n,Q_n)$ are in the same family of solutions and share the same spacetime dynamics, regardless of the value of $\alpha$. A pictorial representation of solutions that are dual to a binary system of electrically charged \glspl{BH} is present in \cref{fig:dual-to-Q-angle}.

These results are valid for any binary system that can have its electromagnetic field, at a given point in time, described by \cref{eq:binary-electromagnetic}. This is the case for different initial data constructions of charged binary systems. Notable examples include multiple charged \glspl{BH} initially at rest~\cite{Alcubierre2009,Zilhao:2013nda,Zilhao2012}, and the merger of a binary of charged \glspl{BH} \cite{Bozzola2019}.

\subsection{Observational consequences}
\label{subsec:observational-consequences}

Even though different solutions can be mapped to each other, it is still possible for an observer to distinguish between dual configurations by looking at electromagnetic observables.

To see how, let us take an observer that is located far away from a source of electromagnetic radiation. Assuming a plane-wave solution that is propagating along the $z$-axis of the form
\begin{subequations}
\label{eq:maxwell-ansatz}
\begin{align}
    \vec{E} &= (E^x,  E^y, 0) \,, \\
    \vec{B} &= (-E^y, E^x, 0) \,,
\end{align}
\end{subequations}
where the components are given by
\begin{subequations}
\begin{align}
    E^x &= A_x \cos \left( k z - \omega t \right)         \,, \\
    E^y &= A_y \cos \left( k z - \omega t + \beta \right) \,,
\end{align}
\end{subequations}
with $A_x$ and $A_y$ the amplitude of the $x$ and $y$ component respectively, $\omega$ the wave frequency, $k = \omega$ the wave number and $\beta$ the phase difference between the $x$ and $y$ components.

For a solution of the form of \cref{eq:maxwell-ansatz}, straightforward computations shows that the duality \cref{eq:duality-fields} reduce to
\begin{subequations}
\begin{align}
    \hat{\vec{E}} &= \begin{pmatrix} \cos\alpha & -\sin\alpha & 0 \\ \sin\alpha & \cos\alpha & 0 \\ 0 & 0 & 1 \end{pmatrix} \vec{E} \,, \\
    \hat{\vec{B}} &= \begin{pmatrix} \cos\alpha & -\sin\alpha & 0 \\ \sin\alpha & \cos\alpha & 0 \\ 0 & 0 & 1 \end{pmatrix} \vec{B} \,,
\end{align}
\end{subequations}
implying that a rotation by angle $\alpha$ in dual space correspond to a physical rotation of the observer by an angle $\alpha$ in the $xy$ plane.

Therefore, any two systems that are within the same family of solutions, will emit electromagnetic radiation that is physically rotated by an angle $\alpha$ around the propagation axis, with respect to each other. For linearly polarized waves, the angle $\alpha$ corresponds to the polarization angle. In the following we will confirm these expectations with full nonlinear evolutions.

\section{Dynamical Spacetimes}
\label{sec:evolutions}

\begin{figure}[t!]
  \centering
  \includegraphics[width=0.88\linewidth]{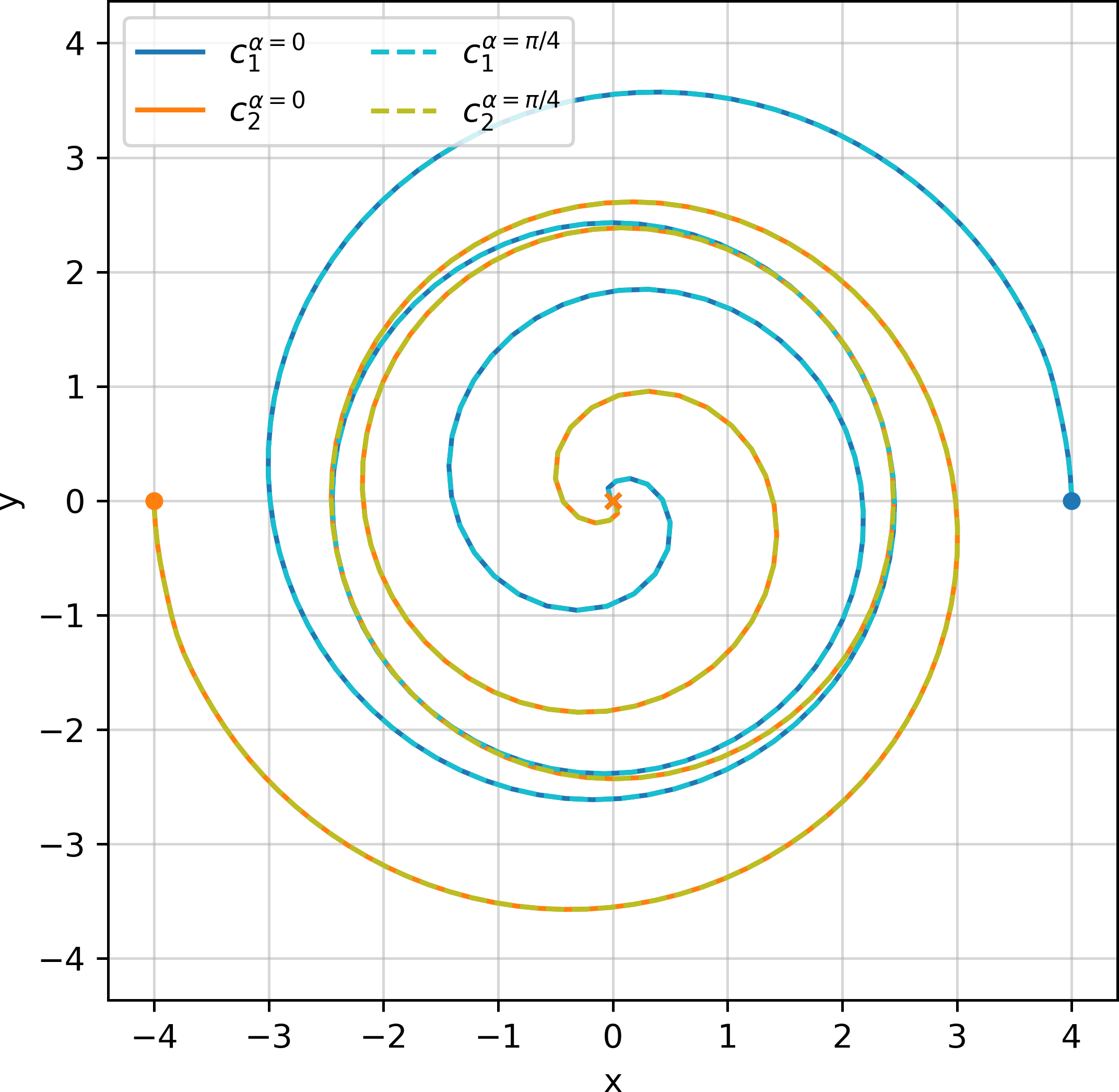}
  \caption{Puncture location of the binary system ($c_1$ and $c_2$) for the electrically charged \glspl{BH} ($\alpha = 0$) in solid lines and the dyonic case $(\alpha = \pi/4)$ in dashed. There is a perfect overlap between both cases.}
  \label{fig:puncture}
\end{figure}

We consider the merger of an equal mass $(M=M_1=M_2=1/2)$ quasi-circular binary for three distinct duality angles: electrically charged $(\alpha = 0)$, dyonic $(\alpha = \pi/4)$ and magnetic $(\alpha = \pi/2)$. The \glspl{BH} are endowed with charge $Q = Q_1 = Q_2 = 1/4$. The evolutions are performed in units where the total mass of the system is set to $M = 1$. For details on the construction of the initial data, see \cref{app:initial-data}.

The evolutions are performed in full 3+1, without imposing any spacetime symmetry. For a review of this formalism, see \cref{app:3+1}. By leveraging the electromagnetic duality, we avoid the need to reformulate the equations of motion and rely on the same numerical infrastructure used in previous studies. We describe the numerical infrastructure in \cref{app:infrastructure} and present detailed convergence studies in \cref{app:convergence}.

The time evolutions show that the spacetime dynamics of the three cases is identical, with the same merger time and final remnant. We can see this in \cref{fig:puncture}, where we plot the puncture location of the binary system for the electrically charged and dyonic cases. The perfect overlap confirms the same spacetime dynamics, as expected from the electromagnetic duality. The magnetically charged case ($\alpha = \pi/2$) also overlaps and is omitted for clarity.

\begin{figure}[t!]
  \centering
  \includegraphics[width=0.9375\linewidth]{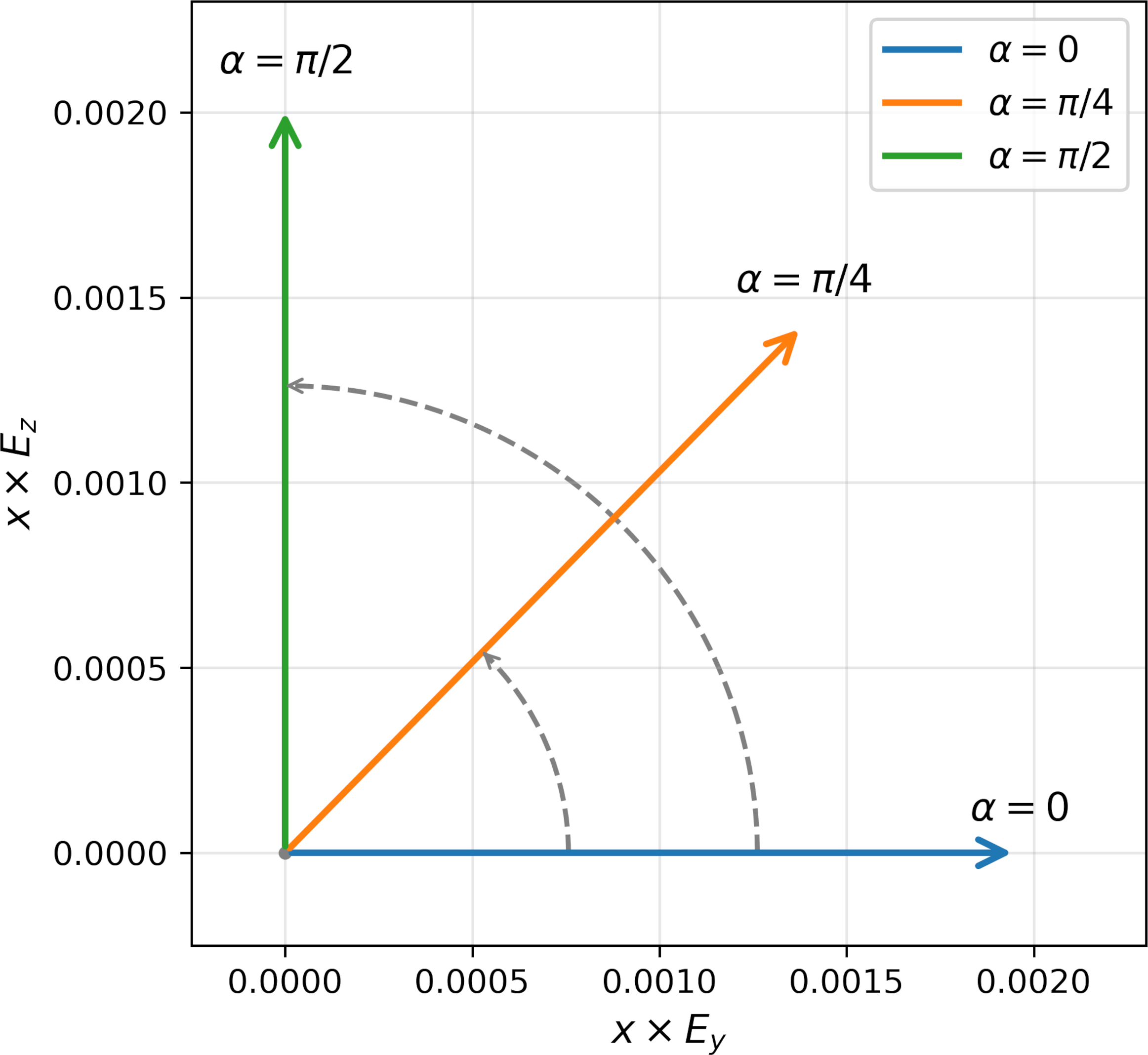}
  \caption{Transverse components of the electric field extracted at $x = 201$ and $t = 240$, for electric ($\alpha = 0$), dyonic ($\alpha = \pi/4$) and magnetic ($\alpha = \pi/2$) configurations. The rotation of the polarization is in numerical agreement with the value of $\alpha$.}
    \label{fig:E}
\end{figure}

\begin{figure*}[t]
  \centering
  \includegraphics[width=0.9\textwidth,height=0.285\textheight]{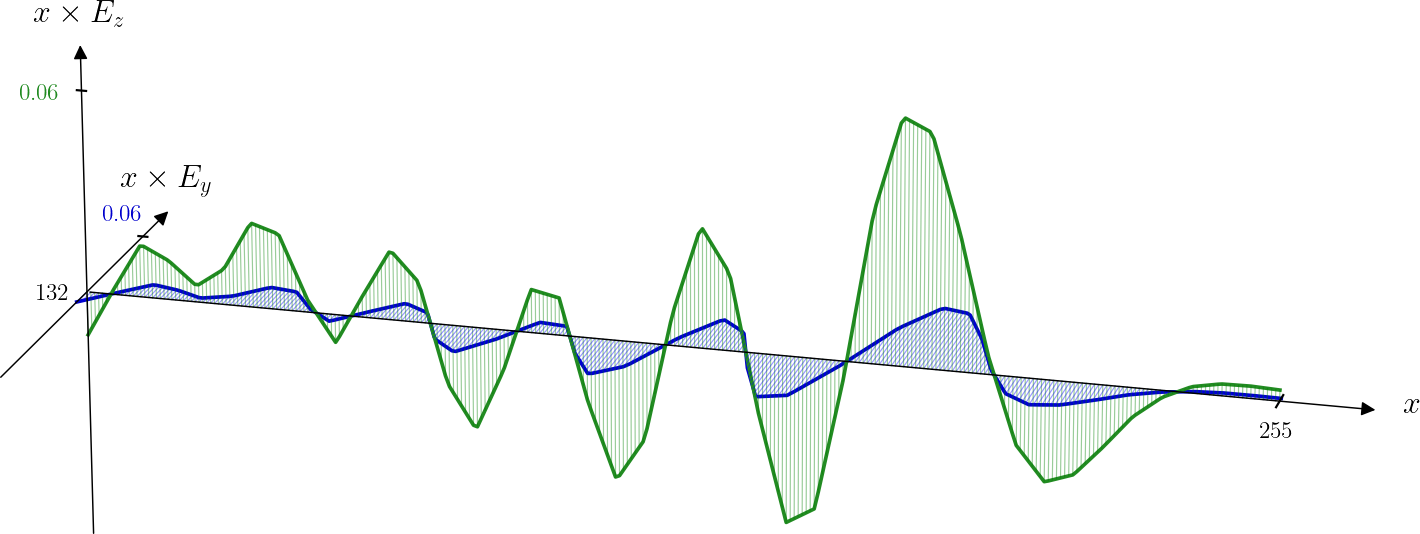}
  \caption{Spatial profile of the electric field extracted between $x \in (132, 255)$ at $t=240$ for purely charged case $(\alpha=0)$ in green and for the purely magnetic case $(\alpha=\pi/2)$ in blue. The waveform for the magnetized solution is rotated by an angle $\alpha = \pi/2$ with respect to the purely electric system. The jagged edges are due to the spacetime discretization.}
  \label{fig:waveform}
\end{figure*}

As for the electromagnetic field, at a large enough distance from the source, we can extract the electromagnetic radiation emitted during the merger. Considering an observer located along the $x$-axis, we measure the orthogonal components of the electric and magnetic fields, $(E_y,E_z)$ and $(B_y,B_z)$ respectively. In \cref{fig:E}, we focus on the orthogonal components of the electric field for the three values of $\alpha$ considered, extracted at $R_\text{ex}=201$. In \cref{fig:waveform}, we present the electric field for the purely electric and purely magnetic case along $x$ at $t=240$. The magnetic field is rotated the same way as the electric field.

From both \cref{fig:E,fig:waveform}, we can see that the polarization of the electromagnetic wave emitted during the merger is rotated by an angle $\alpha$ with respect to each other. This is a direct consequence of the electromagnetic duality, which rotates both the electric and magnetic fields by the same angle $\alpha$, for solutions within the same family. As expected, we can see that the electric field and the magnetic fields, for the same value of $\alpha$, are perpendicular to each other. However, this difference does not extend to other quantities that depend on the electromagnetic field, namely the energy emitted via electromagnetic radiation, which is the same for solutions within the same dual family. We also note that in the present equal-mass, equal-charge configuration, no electromagnetic recoil (kick) is produced; duality predicts that, in asymmetric configurations, any such kick would equally hold under duality rotations.

\subsection{Breaking the duality with gravitational waves}

Although the polarization of the electromagnetic waves carries information about the electromagnetic sector of the system, it is not sufficient, on its own, to distinguish between different members of a duality family. This is because a duality rotation of the electromagnetic field has the same observational effect as a rotation of the observer around the direction of propagation of the wave. Consequently, electromagnetic observations alone are insensitive to the duality angle.

\Glspl{GW}, being unaffected by electromagnetic duality, can provide a geometrical reference frame against which the polarization of the electromagnetic radiation may be compared. However, as we will now show, due to a rotation symmetry of the \glspl{GW} this frame cannot be uniquely determined.

\begin{figure}[b]
  \centering
  \includegraphics[width=0.9\linewidth]{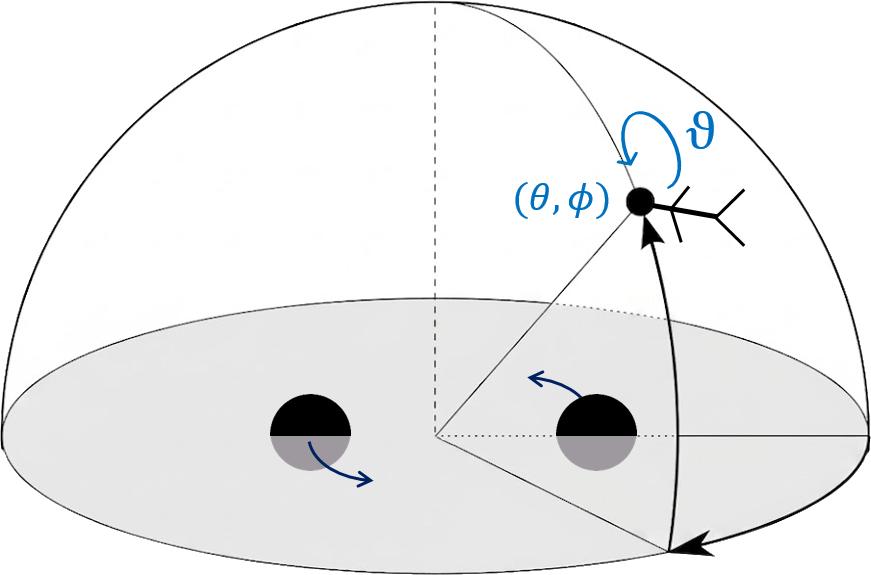}
  \caption{Schematic representation an observer with a sky location of $(\theta, \phi)$, with respect to the binary center of mass, rotating by an angle $\vartheta$ around itself.}
  \label{fig:rotation}
\end{figure}

To understand why, let us take the center of mass of the binary as the origin of the reference frame and assume an observer located very far away on the celestial sphere, at a position $(\theta, \phi)$ in the sky. If the observer is to rotate around itself by angle $\vartheta$ while keeping its sky position $(\theta, \phi)$ fixed, as depicted in \cref{fig:rotation}, then the polarizations of the \glspl{GW} transform as
\begin{subequations}
\begin{align}
  h_+'      &=  h_+ \cos(2 \vartheta) + h_\times \sin(2 \vartheta) \,,  \\
  h_\times' &= -h_+ \sin(2 \vartheta) + h_\times \cos(2 \vartheta) \,.
\end{align}
\end{subequations}

The previous expression shows that a rotation of angle $\vartheta = \pi$ rotates both $h_+$ and $h_\times$ in such a way as to leave them invariant, i.e.:
\begin{equation}
  \vartheta = \pi \implies
    \begin{cases}
      h_+'      = h_+ \\
      h_\times' = h_\times
    \end{cases}
\end{equation}

This implies that, even if the \glspl{GW} can uniquely determine the sky position of the observer, it can also determine its direction with respect to the binary but not its orientation.

Within this partially fixed reference frame, an observer can obtain the duality parameter $\alpha$ by measuring the rotation of the electromagnetic wave relative to its referential. Nevertheless, due to the degeneracy in the orientation of the reference frame of the observer, this reconstruction is only possible modulo $\pi$, i.e., $\alpha \sim \alpha + \pi$.

In particular, this implies that no observer can distinguish between a merger with $\alpha = 0$ (two positive electric charges) from $\alpha = \pi$ (two negative electric charges). The same is true for $\alpha = \pi/2$ (two negative magnetic charges) from $\alpha = 3\pi/2$ (two positive magnetic charges).

More generically, this means that we can use \glspl{GW} to setup a reference frame such that we can tell the nature of the charges (e.g., purely electric, dyonic or purely magnetic), but the rotation symmetry implies that we cannot tell the sign of the charges (whether they are positive or negative).

We stress that additional degeneracies may contribute to the inability to distinguish between dual configurations, depending on the specific configuration of the system and the observer. This includes possible astrophysical effects in the propagation of the radiation, such as Faraday rotation.

\section{Final Remarks}
\label{sec:final-remarks}

In this work we investigated the role of the electromagnetic duality in fully dynamical spacetimes by performing nonlinear numerical evolutions of binary \gls{BH} mergers in the \gls{EM} theory. Starting from a binary of electrically charged \glspl{BH}, we generated a continuous family of configurations through duality rotations of the electromagnetic field, encompassing purely electric, dyonic, and purely magnetic systems.

Our results confirm that electromagnetic duality naturally manifests in fully dynamical \gls{EM} spacetimes. Because the stress-energy tensor is invariant under the duality transformation, all members of the dual family produce identical spacetime dynamics. In our simulations this manifests itself in indistinguishable \gls{BH} trajectories, merger times, and remnant properties across configurations with different electric--magnetic charges.

At the same time, the electromagnetic radiation carries a clear imprint of the duality transformation, with the polarization of the outgoing radiation rotated by an angle equal to the duality parameter. However, this rotation is observationally degenerate with a rotation of the observer around the propagation direction, so electromagnetic measurements alone cannot identify the duality angle. By combining electromagnetic observations with \glspl{GW}, which are unaffected by the duality, one can construct a reference frame to compare against the electromagnetic radiation. Nevertheless, since \glspl{GW} can only determine the axis of the angular momentum and not its direction, the duality parameter is determined only up to $\alpha$ or $\alpha + \pi$. In practice, this means that we can identify the nature of the charges, whether they are purely electric, dyonic, or purely magnetic, but not their sign.

Beyond its conceptual implications, electromagnetic duality also has practical consequences for numerical relativity. Any numerical construction of electrically charged initial data can be straightforwardly extended to produce dyonic or magnetically charged configurations through a simple duality rotation of the electromagnetic fields, without modifying the equations of motion or the numerical infrastructure. This provides a convenient method to explore a broader class of solutions within existing \gls{EM} simulations.

More broadly, our results illustrate how symmetries of the field equations can organize the space of dynamical solutions in general relativity. Extending this analysis to other configurations -- such as unequal-charge binaries, spinning \glspl{BH}, or systems embedded in external electromagnetic environments -- may further clarify the observational and theoretical implications of electromagnetic duality in strong-gravity regimes.

\begin{acknowledgments}

This work is supported by the Center for Research and Development in Mathematics and Applications (CIDMA) (\url{https://ror.org/05pm2mw36}) under the Portuguese Foundation for Science and Technology 
(FCT -- Fundaç\~ao para a Ci\^encia e a Tecnologia, \url{https://ror.org/00snfqn58}), Grants UID/04106/2025 (\url{https://doi.org/10.54499/UID/04106/2025}) and UID/PRR/04106/2025 (\url{https://doi.org/10.54499/UID/PRR/04106/2025}), as well as the projects: Horizon Europe staff exchange (SE) programme HORIZON-MSCA2021-SE-01 Grant No.\ NewFunFiCO-101086251 and  2022.04560.PTDC (\url{https://doi.org/10.54499/2022.04560.PTDC}).
J.F.\ is funded by FCT through project 2023.04333.BD (\url{https://doi.org/10.54499/2023.04333.BD}).
The authors thankfully acknowledge computational resources from RES provided by BSC (MareNostrum) through projects FI-2024-2-0012, FI-2024-3-0007, POR021PROD, and by IFCA (Altamira) through project FI-2025-1-0011.
Computational resources were also provided via FCT through projects 2025.09498.CPCA.A3, 2024.07872.CPCA.A2 (DOI: 10.54499/2024.07872.CPCA.A2 \url{https://doi.org/10.54499/2024.07872.CPCA.A2}) at Deucalion supercomputer, jointly funded by EuroHPC JU and Portugal, and at MareNostrum through project 2024.07059.CPCA.A3 (DOI: 10.54499/2024.07059.CPCA.A3 \url{https://doi.org/10.54499/2024.07059.CPCA.A3}).
This work was supported, in part, by NSF Grant PHY-2145421 and NASA Grants 80NSSC24K0771, 80NSSC26K0343 and ACCESS award PHY190020 to the University of Arizona.

\end{acknowledgments}

\vspace{-5mm}  

\appendix
\section{3+1 Formalism}
\label{app:3+1}

The 3+1 formalism provides the tools for decomposing the four-dimensional field equations into a set of constraints and evolution equations. Here, we give a brief review of this formalism. For a thorough treatment see, e.g., \cite{Gourgoulhon2007,Baumgarte2010,Alcubierre2008}.

Spacetime is foliated by a family of spacelike hypersurfaces $\Sigma_t$, parameterized by a coordinate time $t$. Introducing the lapse function $\alpha$ and the shift vector $\beta^i$, the spacetime line element takes the form
\begin{equation}
    ds^2 = -\alpha^2 dt^2 + \gamma_{ij}\left(dx^i + \beta^i dt\right)\left(dx^j + \beta^j dt\right) \,,
\end{equation}
where $\gamma_{ij}$ is the induced (spatial) metric on $\Sigma_t$. The future-pointing unit normal to $\Sigma_t$ is $n^\mu = \alpha^{-1}(1,-\beta^i)$.

The extrinsic curvature $K_{ij}$ measures how $\Sigma_t$ is embedded in the spacetime manifold,
\begin{equation}
    K_{ij} = -\frac{1}{2\alpha}\left(\partial_t\gamma_{ij} - D_i\beta_j - D_j\beta_i\right) \,,
\end{equation}
where $D_i$ is the covariant derivative compatible with $\gamma_{ij}$.

Projecting Einstein's equation, \cref{eq:EFE}, perpendicular and tangential to $\Sigma_t$ yields the Hamiltonian constraint,
\begin{equation}
    \mathcal{H} \equiv \mathcal{R} + K^2 - K_{ij}K^{ij} - 16\pi\rho = 0 \,,
\end{equation}
the momentum constraints,
\begin{equation}
    \mathcal{M}^i \equiv D_j\left(K^{ij} - \gamma^{ij}K\right) - 8\pi j^i = 0 \,,
\end{equation}
and the evolution equations
\begin{subequations}
\begin{align}
    \partial_t\gamma_{ij} &= -2\alpha K_{ij} + D_i\beta_j + D_j\beta_i \,, \\
    \partial_t K_{ij}     &= \alpha\!\left(\mathcal{R}_{ij} + KK_{ij} - 2K_{ik}K^k{}_j\right) - D_iD_j\alpha \nonumber\\
                          &\quad + \beta^k D_k K_{ij} + K_{ik}D_j\beta^k + K_{jk}D_i\beta^k \nonumber\\
                          &\quad - 8\pi \alpha\!\left(S_{ij} - \tfrac{1}{2}\gamma_{ij}(S - \rho)\right) \,,
\end{align}
\end{subequations}
where $\mathcal{R}_{ij}$ and $\mathcal{R}$ are the Ricci tensor and scalar of $\gamma_{ij}$, and $K \equiv \gamma^{ij}K_{ij}$ is the trace of the extrinsic curvature.

The matter sources measured by the Eulerian observer are the energy density $\rho \equiv T_{\mu\nu}n^\mu n^\nu$, the momentum flux $j^i \equiv - \gamma^i{}_\nu T^{\mu \nu}n_\mu$, the spatial stress tensor $S_{ij} \equiv T_{\mu\nu}\gamma^\mu{}_i\gamma^\nu{}_j$, and its trace $S \equiv \gamma^{ij}S_{ij}$. The source terms are given by
\begin{subequations}
\begin{align}
    \rho   &= \frac{1}{8\pi} \left( E^2 + B^2 \right) \,, \\
    j_i    &= \epsilon_{ijk} E^j B^k \,, \\
    S_{ij} &= \frac{1}{4\pi} \left[ - E_i E_j - B_i B_j + \frac{1}{2} \gamma_{ij} \left( E^2 + B^2 \right)  \right] \,.
\end{align}
\end{subequations}

These are known as the \gls{ADM} equations. However, they are not suitable for numerical integration. Instead, we use the \gls{BSSN} formulation of the Einstein equations \cite{Shibata1995,Baumgarte1998}, that have been shown to be stable for numerical evolutions.

Instead of considering the standard versions of Maxwell equations \cref{eq:EoM-Fmunu}, we will instead consider the enlarged system \cite{Palenzuela2008,Komissarov2007}
\begin{subequations}
\label{eq:EoM-Fmunu-enlarged}
\begin{align}
    \nabla_\mu \left( F^{\mu\nu} + g^{\mu\nu} \Psi \right) &= - \kappa n^\nu \Psi \,, \\
    \nabla_\mu \left( {}^{\star}\!F^{\mu\nu} + g^{\mu\nu} \Phi \right) &= - \kappa n^\nu \Phi \,,
\end{align}
\end{subequations}
where we introduce the scalar $\Psi$, the pseudo-scalar $\Phi$ and the (positive) constant $\kappa$ to improve the control of the constraints. We recover the original system when $\Psi = \Phi = 0$, and set the value of $\kappa$ empirically to obtain adequate convergence.

Considering the decomposition for the electric and magnetic field given in \cref{eq:Decomposition-Fmunu}, we can write the 3+1 version of \cref{eq:EoM-Fmunu-enlarged} as \cite{Moesta2009}
\begin{subequations}
\begin{align}
    &\mathcal{D}_t E^i - \epsilon^{ijk} \partial_j \left( \alpha B_k \right) + \alpha \gamma^{ij} \partial_j \Psi = \alpha K E^i \,, \\
    &\mathcal{D}_t B^i + \epsilon^{ijk} \partial_j \left( \alpha E_k \right) + \alpha \gamma^{ij} \partial_j \Phi = \alpha K B^i \,, \\
    &\mathcal{D}_t \Psi + \alpha \nabla_i E^i = - \alpha \kappa \Psi \,, \\
    &\mathcal{D}_t \Phi + \alpha \nabla_i B^i = - \alpha \kappa \Phi \,,
\end{align}
\end{subequations}
where $\mathcal{D}_t \equiv \partial_t - \mathcal{L}_\beta$, with $\mathcal{L}_\beta$ denoting the Lie derivative along the shift $\beta^i$. For numerical stability, we re-write the previous equations in a \gls{BSSN} form, as presented in~\cite{Zilhao2012}.

To ensure the accuracy of the numerical evolutions, we monitor the electromagnetic constraints
\begin{subequations}
\label{eq:maxwell-constraints}
\begin{align}
    \mathcal{E} \equiv D_i E^i \,, \\
    \mathcal{B} \equiv D_i B^i \,,
\end{align}
\end{subequations}
that should approach zero as the resolution of the simulation increase.

\section{Initial Data}
\label{app:initial-data}

We summarize the construction of initial data for a multiple electrically charged BHs using the TwoChargedPunctures code \cite{Bozzola2019}, that was based on the original \texttt{TwoPunctures} code \cite{Ansorg2004}. The formalism is based on the \gls{CTT} decomposition with the puncture approach.

Constraining ourselves to the case of $N=2$ electrically charged \glspl{BH}, and initially vanishing magnetic field, the gravitational constraint equations become
\begin{subequations}
\begin{align}
  \mathcal{R} + K^2 - K_{ij}K^{ij} - 2 E^2 &= 0 \,, \\
  D_j\left(K^{ij} - \gamma^{ij}K\right) &= 0 \,.
\end{align}
\end{subequations}

We adopt a conformally flat spatial metric and maximal slicing,
\begin{equation}
  \gamma_{ij} = \psi^4 \eta_{ij} \,, \quad K = 0 \,,
\end{equation}
where $\psi$ is the conformal factor and $\eta_{ij}$ the conformally flat metric. Under these conditions, the extrinsic curvature reads $K_{ij} = \psi^{-2} \bar{A}_{ij}$. Suppressing the radiative degrees of freedom of $\bar{A}_{ij}$, we can write the conformal traceless extrinsic curvature as
\begin{equation}
  \bar{A}_{ij} = 2V_{(i,j)} - \tfrac{2}{3}\delta_{ij}\partial_k V^k \,,
\end{equation}
where we introduced the vector $V^i$. Considering the Bowen-York solution for the gravitational sector \cite{Brandt1997,Ansorg2004}
\begin{equation}
  V^i = \sum_{n} \left( - \frac{7}{4} \frac{P^i_n}{R_n} - \frac{1}{4} \delta_{jk} x_n^jP_n^k \frac{x_n^i}{R_n^3} + \frac{\bar{\epsilon}^i_{\,\,jk} x_{n}^jS_{n}^k}{R_n^3} \right) \,,
\end{equation}
where $P_n^i$ and $S_n^i$ are the linear and angular momenta of the $n$-th \gls{BH}, located at $x_n$, and $R_n = |x-x_n|$. This ansatz, corresponding to the initial data of several black holes in vacuum with arbitrary momenta and spins, automatically satisfies the momentum constraint.

The conformal factor is decomposed as
\begin{equation}
  \psi = \sqrt{\kappa^2 - \varphi^2} \,,
\end{equation}
with
\begin{equation}
  \kappa = 1 + u + \sum_n \frac{M_n}{2R_n} \,, \quad \varphi = \sum_n \frac{Q_n}{2R_n} \,,
\end{equation}
where $u$ is a (smooth) correction that transforms Hamiltonian constraint into the elliptic equation
\begin{multline}
  \label{eq:ham-constraint-u}
  \kappa \Delta u + \partial_a\kappa\,\partial^a\kappa - \partial_a\varphi\,\partial^a\varphi - \partial_a\psi\,\partial^a\psi \\
    + \frac{1}{8}\psi^{-6}\bar{A}_{ij}\bar{A}^{ij} + 2\pi\psi^{-2}\bar{\rho} = 0 \,,
\end{multline}
where $\Delta$ is the flat-space Laplacian and $\bar{\rho} \equiv \psi^8 \rho$. This equation is solved numerically using a modified version of the \texttt{TwoPunctures} \cite{Ansorg2004} code adapted for charged black holes.

For the electromagnetic sector, introducing the conformal electromagnetic fields $\bar{E}^i = \psi^6 E^i$ and $\bar{B}^i = \psi^6 B^i$, the Maxwell constraints \cref{eq:maxwell-constraints} decouple from $\psi$ and reduce to
\begin{subequations}
\begin{align}
  \partial_i \bar{E}^i &= 0  \,, \\
  \partial_i \bar{B}^i &= 0 \,.
\end{align}
\end{subequations}
Their linearity allows for superposition. Endowing each \gls{BH} with a point charge electromagnetic field,
\begin{equation}
  \bar{E}^i = \sum_n \frac{Q_n}{R_n^2}\,\hat{R}^i_n \,,
\end{equation}
the physical fields follow from $E^i = \psi^{-6}\bar{E}^i$.

Although the code is more general, we restrict ourselves to equal mass, non-spinning and quasi-circular binaries. The dyonic and magnetic configurations are then obtained by applying the duality rotation \cref{eq:duality-fields} to the initial data.

\vspace{-5mm}  

\section{Numerical Infrastructure}
\label{app:infrastructure}

We perform time evolutions using the Einstein Toolkit~\cite{ETK,Zilhao:2013hia} with the \texttt{EMG} thorn \cite{Zilhao2012}, the \texttt{Carpet} driver for mesh refinement \cite{Carpet}, \texttt{AHFinderDirect} for horizon tracking \cite{AHFinderDirect}, and BSSN metric evolution \cite{Baumgarte1998,Shibata1995} via the \texttt{LeanBSSNMoL} thorn~\cite{Canuda,Sperhake:2006cy}. The initial momenta is generated by \texttt{NRPyPN} \cite{NRPyPN} using post-Newtonian approximations.

Simulations run on a $252 \times 252 \times 252$ grid, with 3 additional ghost points and without symmetries, using radiative boundary conditions and eight-level box-in-a-box mesh refinement with base grid spacing $h = 3$ on the coarsest mesh. The mesh automatically follows the puncture trajectory during evolution. The code uses fourth-order spatial finite differences and Kreiss-Oliger dissipation (continuous scheme with $\epsilon = 0.3$ on the finest level) \cite{Bozzola2021}. Time integration is performed with a fourth-order Runge-Kutta.

Post processing is performed with Kuibit \cite{Kuibit} and Visit~\cite{Visit}. Analysis code and results are available in \cite{dual-collision}.

\section{Gravitational wave extraction}
\label{app:gravitational-waves}

To characterize gravitational radiation in numerical relativity, we make use of the Newman-Penrose formalism \cite{Newman1962}. In this approach, we introduce the Weyl scalar
\begin{equation}
    \Psi_4 \equiv C_{\alpha\beta\gamma\delta}\, q^\alpha \bar{m}^\beta q^\gamma \bar{m}^\delta
\end{equation}
that encodes outgoing gravitational radiation in the wave zone, where $C_{\alpha\beta\gamma\delta}$ is the Weyl tensor and
\begin{subequations}
\begin{align}
    \ell^\mu &= \frac{1}{\sqrt{2}} \left( n^\mu + r^\mu \right)                     \,, \\
    q^\mu    &= \frac{1}{\sqrt{2}} \left( n^\mu - r^\mu \right)                     \,, \\
    m^\mu    &= \frac{1}{\sqrt{2}} \left( 0, \hat{\varphi} - i \hat{\theta} \right) \,,
\end{align}
\end{subequations}
are the null tetrad vectors, where $n^\mu$ is the unit normal vector from \cref{app:3+1},  $r^\mu \equiv (0, \hat{r})$ and $(\hat{r},\hat{\theta},\hat{\varphi})$ are the unit radial, polar and azimuthal vectors.

At large extraction radius and in an asymptotically flat frame, $\Psi_4$ is related to the two transverse-traceless gravitational-wave polarizations by
\begin{equation}
    \Psi_4 = \ddot h_+ - i \ddot h_\times \equiv \ddot h \,,
\end{equation}
where $h \equiv h_+ - i h_\times$, and overdots denote derivatives with respect to retarded time. Thus, once $\Psi_4$ is extracted, the strain polarizations can be obtained by time integration, with the real part of $\Psi_4$ fully determining the polarization $h_+$ and its imaginary part determining  $h_\times$.

\section{Convergence Studies}
\label{app:convergence}

To assess numerical convergence, we perform higher-resolution counterparts of the simulations mentioned, using a finest mesh spacing of $h=2.5$ on a $250 \times 250 \times 250$ grid, to be compared against the baseline resolution of $h=3$.

\begin{figure}[h!]
    \centering
    \includegraphics[width=\linewidth]{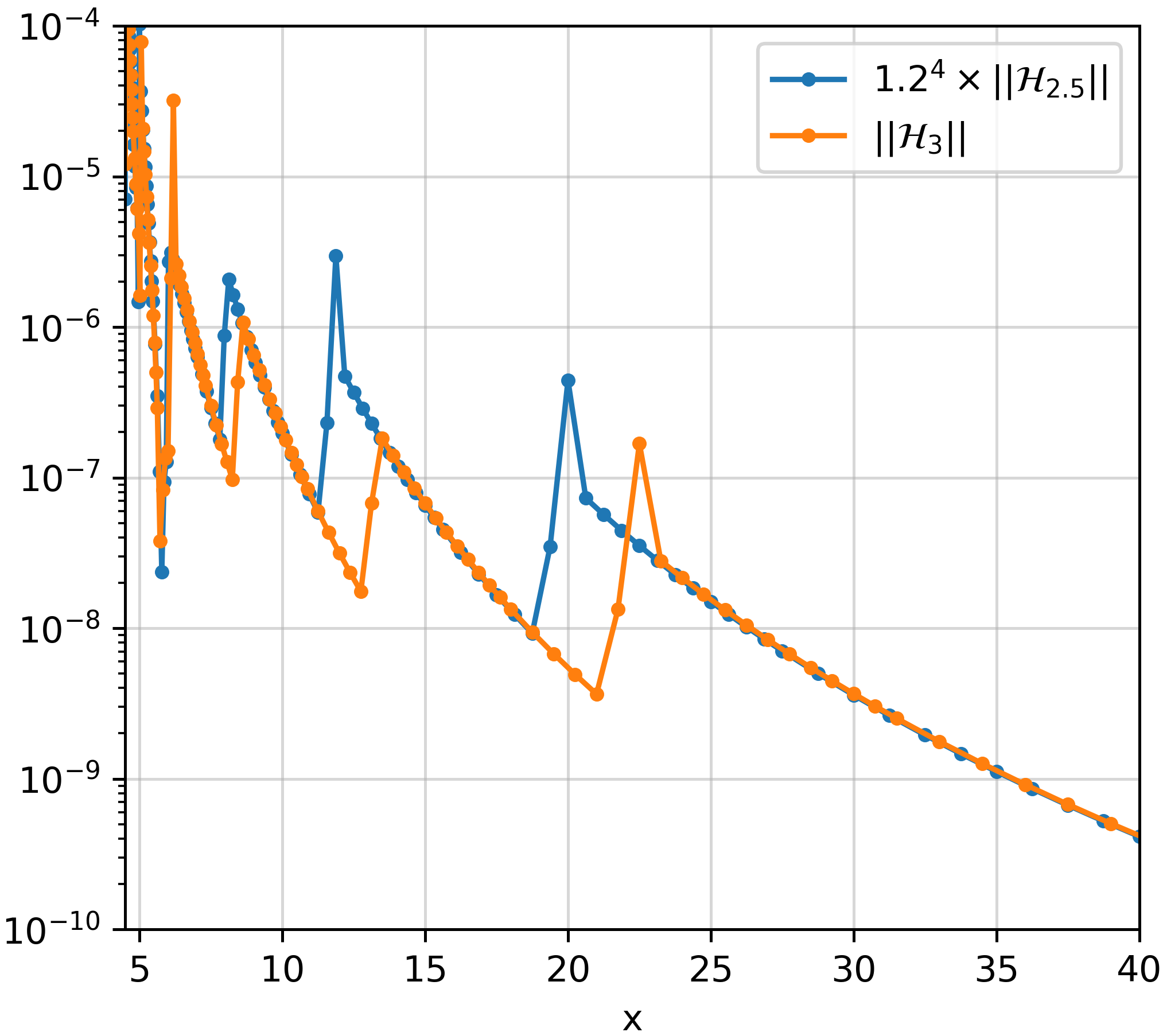}
    \caption{The Hamiltonian constraint $\mathcal{H}$ as a function of $x$ at $t=0$, for two different grid spacings on the coarsest mesh, $h=2.5$ and $h=3$.}
    \label{fig:hc_vs_x}
\end{figure}

\begin{figure}[h!]
    \centering
    \includegraphics[width=\linewidth]{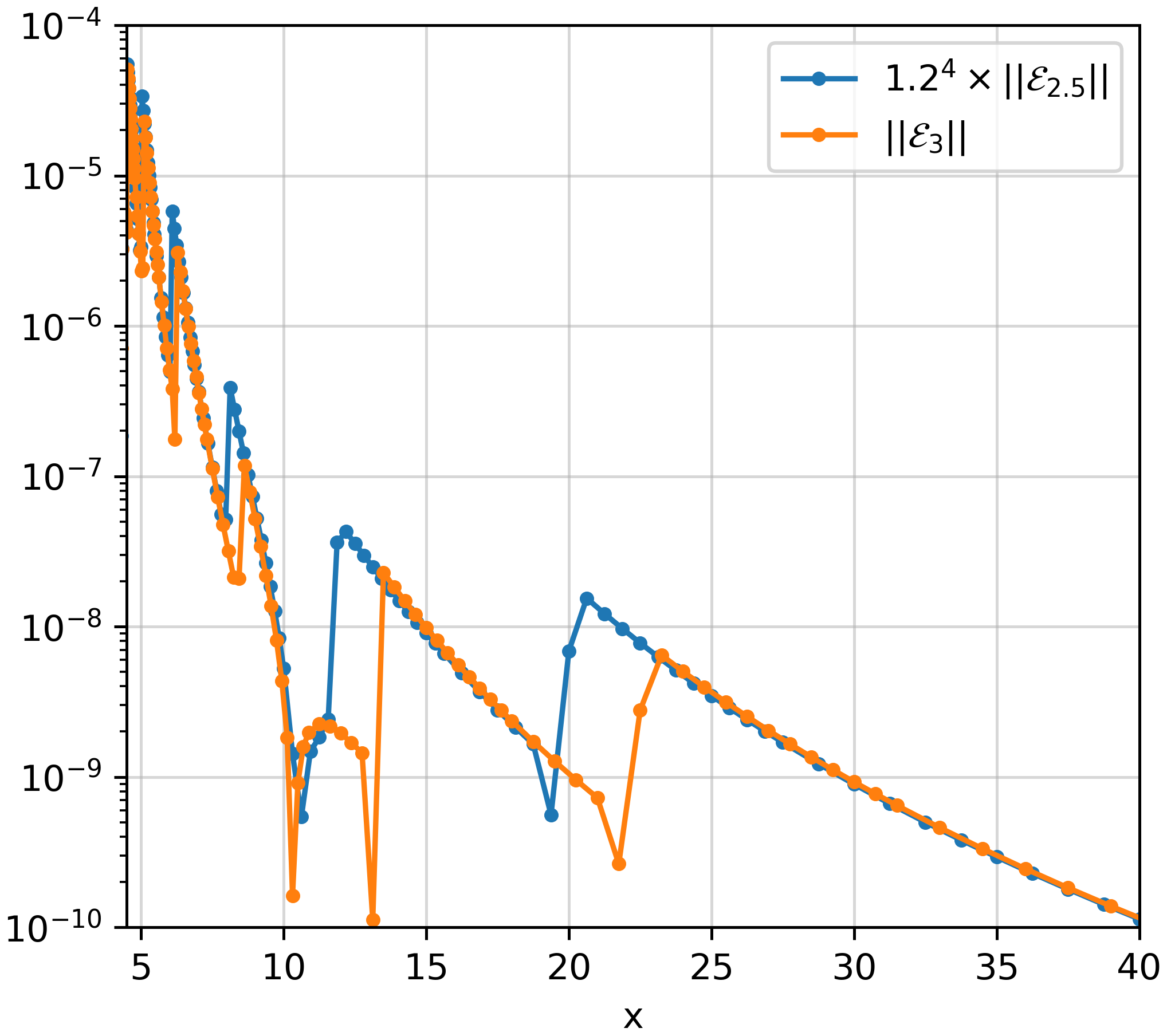}
    \caption{The divergence of the electric field $\mathcal{E}$ as a function of $x$ at $t=0$, for two different grid spacings on the coarsest mesh, $h=2.5$ and $h=3$.}
    \label{fig:Ec_vs_x}
\end{figure}

At the level of the initial data, we observe fourth-order convergence in both the Hamiltonian constraint $\mathcal{H}$ and the divergence of the electric field $\mathcal{E}$, as shown in \cref{fig:hc_vs_x,fig:Ec_vs_x}. The overlapping lines show that the convergence order is consistent with the fourth-order finite difference scheme used in the numerical implementation. The divergence of the magnetic field $\mathcal{B}$ is trivially zero at $t=0$ by construction of the initial data. 

\begin{figure}[h!]
    \centering
    \includegraphics[width=\linewidth]{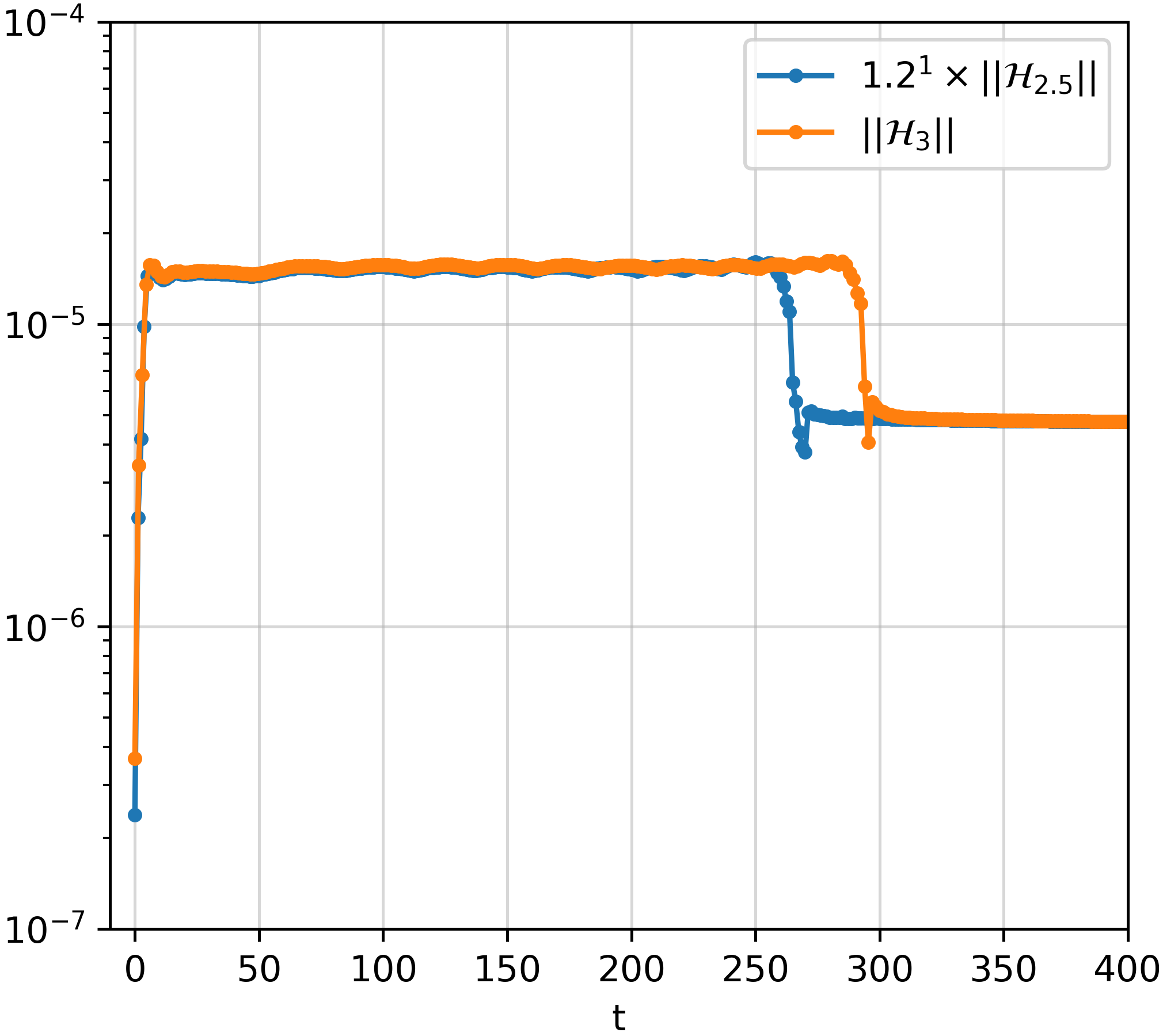}
    \caption{Norm of the Hamiltonian constraint $\mathcal{H}$ as a function of time, for two different grid spacings on the coarsest mesh, $h=2.5$ and $h=3$.}
    \label{fig:hc_norm}
\end{figure}

\begin{figure}[h!]
    \centering
    \includegraphics[width=\linewidth]{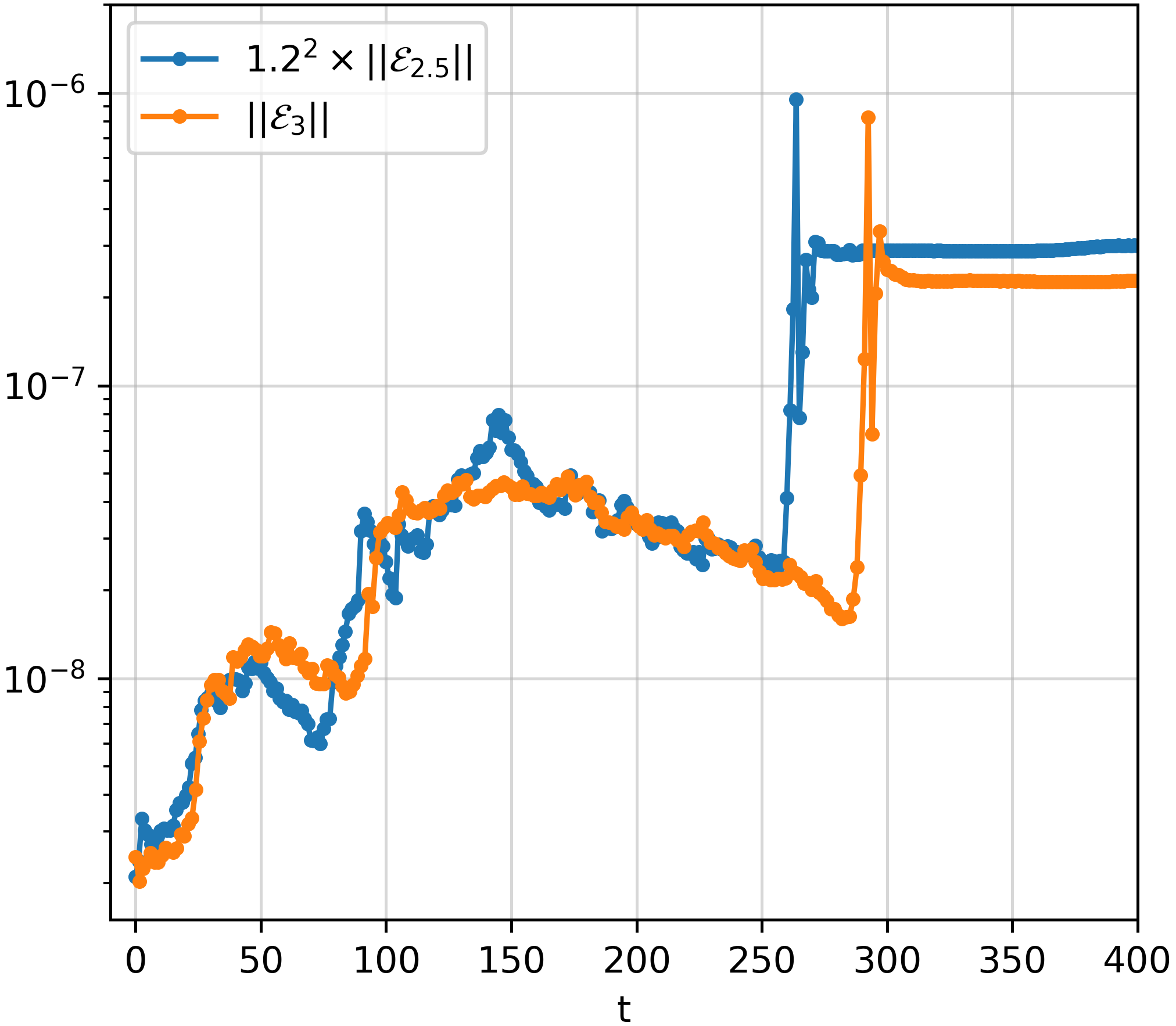}
    \caption{Norm of the divergence of the electric field $\mathcal{E}$ as a function of time, for two different grid spacings on the coarsest mesh, $h=2.5$ and $h=3$.}
    \label{fig:Ec_norm}
\end{figure}

\begin{figure}[h!]
    \centering
    \includegraphics[width=\linewidth]{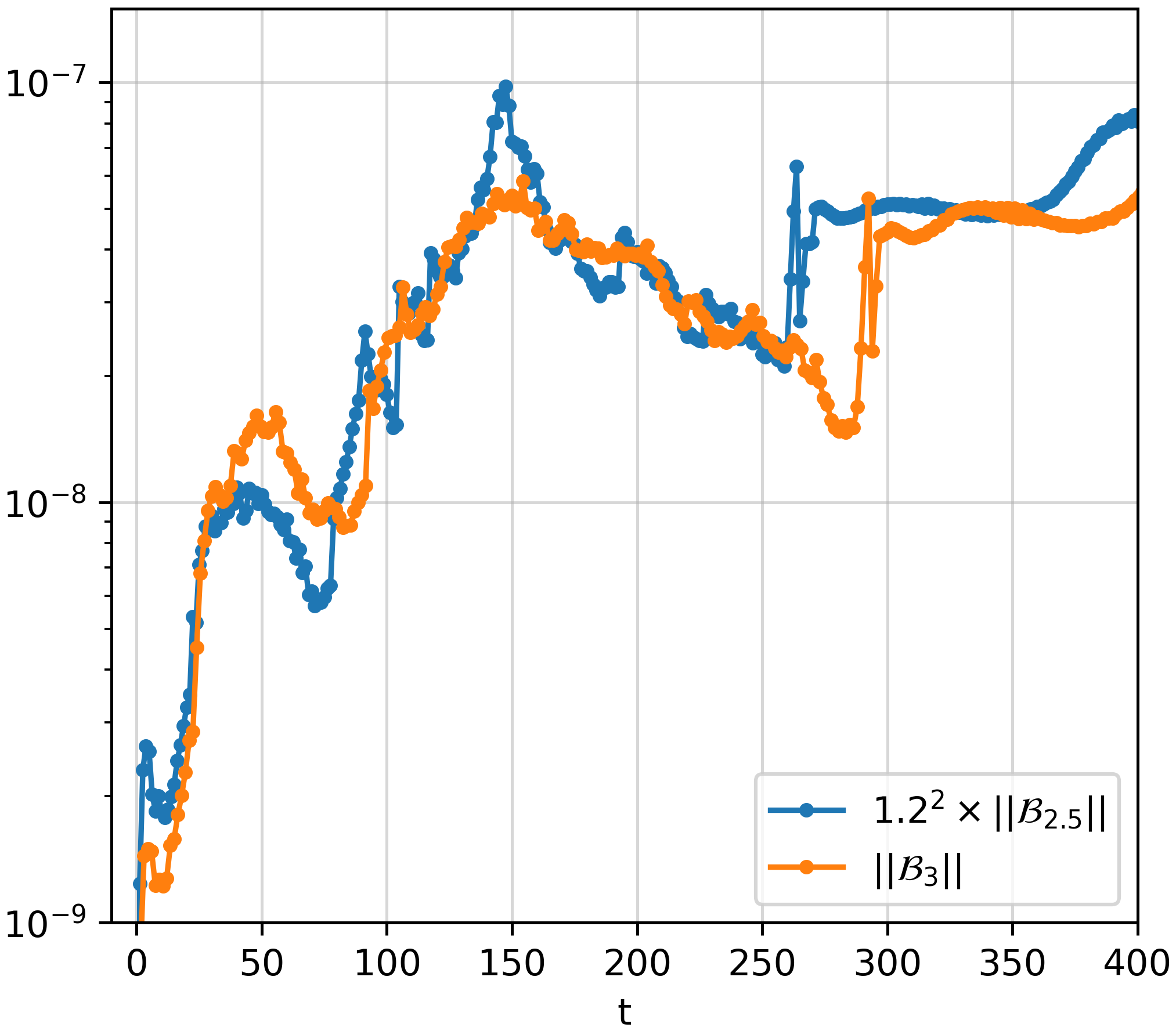}
    \caption{Norm of the divergence of the magnetic field $\mathcal{B}$ as a function of time, for two different grid spacings on the coarsest mesh, $h=2.5$ and $h=3$.}
    \label{fig:Bc_norm}
\end{figure}

During the time evolution, we monitor the norms of the Hamiltonian and Maxwell constraints, shown in \cref{fig:hc_norm,fig:Ec_norm,fig:Bc_norm}. Although the scheme is formally fourth order, the use of second-order operations used in the prolongation operations of \texttt{Carpet} reduces the expected convergence order to second order throughout the evolution. This expected order is achieved for both Maxwell constraints $\mathcal{E}$ and $\mathcal{B}$, while the Hamiltonian constraint $\mathcal{H}$ converges only at first order.

A phase offset in the constraint norms is visible at around $t=250$. This arises because different resolutions produce slightly different merger times, introducing a systematic time shift in quantities extracted from the simulation.

At around $t=500$, a numerical instability develops near the outer boundaries, manifesting as a growth in the electromagnetic field that drives an exponential increase in the constraints and eventually terminates the simulation. The origin of this instability has not been fully identified.

\bibliography{bibliography}

\end{document}